\documentclass[12pt]{article}
\usepackage{graphicx}

\makeatletter

\renewcommand{\section}{\@startsection {section}{1}{\z@}
{-3.5ex plus -1ex minus -.2ex}{2.3ex plus .2ex}{\normalsize\bf}}

\renewcommand{\subsection}{\@startsection{subsection}{2}{\z@}
{-3.25ex plus -1ex minus -.2ex}{1.5ex plus .2ex}{\normalsize\it}}

\def\abstract{\if@twocolumn
\section*{\abstractname}
\else \small
\quotation
\fi}
\def\endabstract{\if@twocolumn\else\endquotation\fi}

\renewcommand{\@makefnmark}{\hbox{\mathsurround=0pt
$^\dagger$}}
\renewcommand{\@makefntext}[1]{\parindent=1em\noindent
\hbox to 1.8em{\hss$^\dagger$}#1}

\def\thebibliography#1{\section*{\refname\@mkboth
 {\uppercase{\refname}}{\uppercase{\refname}}}\list
 {[\arabic{enumi}]}{\settowidth\labelwidth{[#1]}\leftmargin\labelwidth
 \advance\leftmargin\labelsep\parsep=0em\itemsep=0em
 \usecounter{enumi}}
 \def\newblock{\hskip .11em plus .33em minus .07em}
 \sloppy\clubpenalty4000\widowpenalty4000
 \sfcode`\.=1000\relax}

\makeatother

\begin{document}

\begin{center}

{\normalsize\bf
 RESONANT SPIN-DEPENDENT TUNNELING IN SPIN-VALVE JUNCTIONS
 IN THE PRESENCE OF PARAMAGNETIC IMPURITIES. }\\
\bigskip
A. Vedyayev$^{1,2}$, D. Bagrets$^{1,2}$, A. Bagrets$^{2}$,
and B. Dieny$^{1}$
\medskip \\
\small{\it
$^1$CEA/Grenoble, D\'epartement de Recherche Fondamentale sur la
Mati$\grave e$re Condens\'ee,
SP2M/NM, 38054 Grenoble, France \\
$^2$Department of Physics, M.~V.~Lomonosov Moscow State University,
119899 Moscow, Russia }
\end{center}

\begin{abstract}
The tunnel magnetoresistance (TMR) of F/O/F magnetic junctions, (F's are 
ferromagnetic layers and O is an oxide spacer) in the presence of magnetic 
impurities within the barrier, is investigated. We assume that magnetic 
couplings exist both between the spin of impurity and the bulk magnetization of
the neighboring magnetic electrode, and 
between the spin of impurity and the spin of tunneling electron. 
Consequently, the resonance levels of the system formed by a tunneling electron 
and a paramagnetic impurity with spin $S=1$, are a 
sextet. As a result the resonant tunneling depends on the direction of the 
tunneling electron spin. At low temperatures and zero bias voltage the TMR of 
the considered system may be larger than TMR of the same structure without 
paramagnetic impurities. It is calculated that an increase in temperature 
leads to a decrease in the TMR amplitude due to excitation of spin-flip processes 
resulting in mixing of spin up and down channels. It is also shown that 
asymmetry in the location of the impurities within the barrier can lead to 
asymmetry in $I(V)$ characteristics of impurity assisted current and 
two mechanisms responsible for the origin of this effect are established. The 
first one is due to the excitation of spin-flip processes at low voltages and 
the second one arises from the shift of resonant levels inside the insulator 
layer under high applied voltages.
\end{abstract}

\section{Introduction}

The observation of the large tunneling magnetoresistance effect at room
temperature in tunnel junctions of the
form M/O/M' (where M and M' are magnetic metals and O is an oxide tunnel
barrier) has stimulated  a renewed interest for these
systems \cite{Moodera,Miyazaki,Gallagher}. Besides
the fundamental interest for spin-polarized transport,
these structures are also foreseen as potential candidates
for sensitive magnetic sensors and memory cells in random access memory
devices. The first model of spin-dependent tunneling  in the framework
of classical quantum mechanics was proposed by S{\l}onczewski~\cite{Sloncz}.
However, in this approach no scattering of electrons in the magnetic metallic
electrodes was taken into account.  This model has been subsequently
developed in Refs.~\cite{Vedy1997,Vedy1999} by using the Kubo
formalism of linear response. The effects of elastic impurity scattering
inside the metallic layers and at interfaces between the dielectric and
conductive layers could then be incorporated in the model.
On the other hand, it is well known \cite{Lifshitz}
that the presence of impurities inside the potential barrier
can lead to the mechanism of resonant tunneling
when the localized electronic states within the gap of the
insulator formed by embedded atoms lie close to the chemical potential
of the system. This situation was qualitatively studied in mesoscopic
semiconductor system \cite{Larkin} in case of one- and two-impurity resonant
channels by means of classical quantum mechanical treatment. The same approach
afterwards has been  used in \cite{Bratkovsky} with the application to the
impurity-assisted tunneling magnetoresistance (TMR). The numerical analysis
of this problem which was carried out in Refs.~\cite{Tsymbal,Itoh}
should also be mentioned. In paper~\cite{Bratkovsky} only the case of spinless
impurities was considered, and the author came to the conclusion
that the TMR amplitude decreased due to the impurity assisted
tunneling. The problem of the paramagnetic impurity assisted
tunneling in tunnel magnetic junctions was investigated recently
in Ref.~\cite{Guinea}. The author claim, that he investigated
resonant tunneling through impurity resonance level, however
he did not introduce the line-width of this resonance,
which, as it will be shown below, does depend on the
position of the impurity atom inside the barrier as well
on the magnetic configuration of the magnetic layers.
As it will be shown below, namely these line-widths define
the value of the tunneling conductance and the amplitude
of the TMR for spin-conserving and spin-flip resonant
tunneling. An attempt of the analysis the same problem has
also been undertaken in Ref.~\cite{Jansen}, but nevertheless
the microscopic mechanism of electron scattering on the paramagnetic
impurity has not been taken into account.

  In this paper, we propose a renewed study of the problem of
impurity-assisted tunneling in spin-valve
junctions of the form F$_1$/O/F$_3$,
where F's are ferromagnetic electrodes and O is an insulating barrier with
embedded paramagnetic impurities, that incorporates the effect of both
elastic and non-elastic spin-flip scattering due to the exchange
interaction between the itinerant electrons forming the tunneling current
and the localized spins of impurities. It will be shown that non-elastic
scattering has an essential impact not only on the temperature variation
of the TMR (which is a well established result \cite{Levy1997}) but also on the
I--V characteristics of considered structures.
The latter effect
was predicted in Ref.~\cite{Levy1997}, where the TMR dependence
on the electron scattering on interfacial magnons was investigated.

\section{Model}
\subsection{Kubo formula and general expression for
the conductivity of the system}

  The following simplified model is adopted throughout the paper.
First of all, the thickness of an oxide layer is supposed to
be much smaller than its in-plane dimension, so that the
system may be considered as homogeneous in the $xy$--plane
(parallel to the interfaces) and inhomogeneous only in the
$z$--direction (growth direction). Within each layer, the
electrons are described as a free-electron gas and they undergo
scattering on the 3-D $\delta$-function impurity potential
within the insulating barrier. Within these
approximations, the Hamiltonian of the system has the form
\[
  \hat{H} = \hat{H_0} + \hat{H}_{\rm int},
\]
where
\begin{eqnarray}
\label{Hamilt}
  \hat{H}_0 = -\frac{\hbar^2}{2m(z)}\Delta + U(z) -
  2\mu_B H_z^{\rm eff}(z)(\hat{s_z} + \hat{S_z})
\\
 \hat{H}_{\rm int} = \sum_i a_0^3 \delta({\bf r} - {\bf c}_i)
\left\{ \varepsilon_0 - J({\bf sS})  \right\}.
\nonumber
\end{eqnarray}
Here the summation is performed over the location of impurities
${\bf c}_i$~inside the barrier,  $a_0$~ is the lattice constant,
$\varepsilon_0$~denotes the scattering potential amplitude on the
impurity, $J$~is responsible for the {\it s--d\/} type exchange interaction
between a conduction electron spin ${\bf s}$ and the impurity spin ${\bf S}$,
$U(z)$ is a model step-like potential seen by the conduction electron as
it is represented in Fig.~1. We take into account the exchange splitting of
the $d$--band by introducing different values $V_{\mu}^{1,3}$
for the position of the bottom of the conduction band
in F$_1$ and F$_3$, depending on the mutual orientation of magnetization
in the layers and the spin $\mu = \uparrow, \downarrow$
of the conduction electron.
$H^{\rm eff}_z(z)$ represents the effective field
acting on impurity and electron spins inside the barrier. The origin of this
field is the super-exchange between the spins in the bulk of ferromagnetic
layer and in insulating layer. We suppose that $H^{\rm eff}_z(z)$
decreases exponentially with the distance from the interface in the
depth of the oxide layer. $m(z)$ corresponds to the
effective electron mass that we suppose is equal to $m$ in
the ferromagnetic layers and to $m_0$ in the insulator.
Throughout the paper, it is expressed
in units of bare electron mass $m_e$. We also assume that the mass
of free-like electrons in ferromagnet is slightly differs from $m_e$,
i.e. $m\approx 1$ and we will eliminate it from all subsequent
expressions.

\begin{figure}
\begin{center}
\includegraphics[scale=1.0, angle=0]{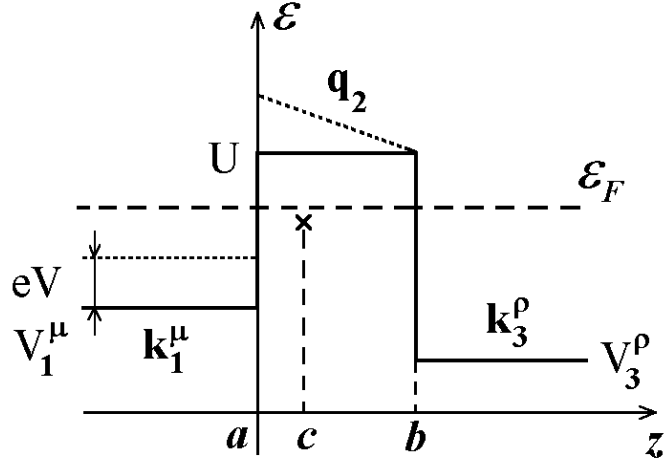}
\end{center}
\caption{\small The potential profile seen by electron propagating through the 
F/O/F
junction comprising impurity defect inside the oxide spacer. $k_1^{\mu}$,
$k_3^{\rho}$, $q_2$ are the momenta inside the magnetic layers and oxide 
barrier,
respectively. $V_{1(3)}^{\mu(\rho)}$ denotes the spin-dependent conduction
band bottom, $U$ is a level of the barrier and $\varepsilon_F$ is Fermi energy.
The paramagnetic impurity is located at point $c$. The variation of the 
potential
profile under high bias voltage is indicated by the dashed line.}
\end{figure}

  We start from the Keldysh technique for Green functions together with
Kubo exact formula of linear response theory for the static
conductivity which relates its real part with
the current-current correlation function and
may be written in the form \cite{Kubo}
\begin{eqnarray}
\sigma_{\mu\rho}({\bf r},{\bf r'}) =
\frac{1}{2k_B T}\int_{-\infty}^{+\infty}
\Bigl\langle
j_{\rho}({\bf r'},t')j_{\mu}({\bf r},t)
\Bigr\rangle\,d(t-t')
\nonumber
\\
 = \frac{1}{2k_B T}\left(\frac{e\hbar}{2m}\right)^2
\int_{-\infty}^{+\infty}\Bigl\langle
G^{<}_{\mu\rho}({\bf r},t,{\bf r'},t')
\stackrel{\leftrightarrow}{\nabla}_{\bf r}
\stackrel{\leftrightarrow}{\nabla}_{\bf r'}
G^{>}_{\rho\mu}({\bf r'},t',{\bf r},t)\Bigr\rangle\,d(t-t'),
\label{Kubof}
\end{eqnarray}
where $\mu$, $\rho$ denote the projections of the spin of the electrons,
$\stackrel{\leftrightarrow}{\nabla}_{\bf r} =
\frac{1}{2}(\stackrel{\rightarrow}{\nabla}_{\bf r} -
\stackrel{\leftarrow}{\nabla}_{\bf r} )$
is the asymmetric gradient operator and $G^{<}_{\mu\rho}$
and $G^{>}_{\rho\mu}$ are corresponding Green functions in
Keldysh formalism \cite{Keldysh}. $\langle\dots\rangle$ represents
the quantum statistical averaging over the distribution of impurities
and degrees of freedom of impurity spin. This expression is most
general and holds both for the elastic impurity and defect
scattering or inelastic, including magnon
and phonon, scattering. To evaluate expression~(\ref{Kubof})
one needs to introduce the retarded Green function
$G^R_{\mu\rho}(z,z',\kappa,\varepsilon)$
that in our case is defined by the following differential equation:
\[
\left\{
\varepsilon + \frac{\hbar^2}{2m}\frac{\partial^2}{\partial z^2} -
\frac{\kappa^2}{2m} - U(z) \right\}
G^R_{\mu\rho}(z, z', \kappa, \varepsilon) =
\delta_{\mu\rho}\delta(z-z')
\]
in the mixed real-space momentum representation \cite{Vedy1997,Vedy1999},
where $\kappa = (\kappa_x,\kappa_y)$ is the component of the electron
momentum in $xy$-plane of the layers and $z$ is the coordinate
perpendicular to $xy$-plane. We should note, that by definition
the conductivity~(\ref{Kubof}) is defined as a linear response on the
externally applied electric field and does not depend on
$z$ and $z'$ because of the obvious condition
$\displaystyle \frac{\partial j(z)}{\partial z} = 0$.

 Let now denote $k_1^{\mu} = \sqrt{2(\varepsilon-V_1^{\mu})}$,
$k_3^{\mu} = \sqrt{2(\varepsilon-V_3^{\mu})}$  momenta
of electrons with energy $\varepsilon$ and spin $\mu$ in the ferromagnetic 
layers and $q_0 = \sqrt{2m_0(U - \varepsilon)}$ is an imaginary momentum
inside the barrier. 
By introducing the following functions on $x = \kappa/q_0$:
\[
 c_1^{\mu}(x) = \sqrt{k_1^{\mu\,2} - q_0^2 x^2}, \quad
 c_3^{\mu}(x) = \sqrt{k_3^{\mu\,2} - q_0^2 x^2}, \quad
 q_2(x) = q_0\sqrt{1 + x^2},
\]
the final expression for the conductance of the system,
comprising only one impurity, located at point~${\bf c}$,
at given temperature $T$ is written as follows:
\[
\sigma(T,{\bf c}) = \sigma_0(T) +
\sigma^{\rm imp}(T,{\bf c}).
\]
The first term is given by
\begin{equation}
\sigma_0(T) = \frac{q_0^2 e^2}{2\pi\hbar}
\sum_{\mu}
\int_{-\infty}^{+\infty}d\varepsilon
\left(-\frac{\partial f(\varepsilon)}{\partial\varepsilon}\right)
\int_0^{x_0^{\mu}} \frac{xdx}{2\pi}
\frac{16 c_1^{\mu} c_3^{\mu} m_0^2 q_2^2 e^{-2q_2w}}
{(m_0^2 c_1^{\mu\,2} + q_2^2)(m_0^2 c_3^{\mu\,2} + q_2^2)},
\label{sigma0}
\end{equation}
where $x_0^{\mu} = \min
\left\{\frac{k_1^{\mu}}{q_0}, \frac{k_3^{\mu}}{q_0}\right\}$,
$f(\varepsilon) =
\left[1 + e^{\beta(\varepsilon -
\varepsilon_F)} \right]^{-1}$
is Fermi function, and $w=b-a$ is the width of the insulating spacer.
It represents the well known result for the pure tunneling conductance
\cite{Sloncz}. The second term
$\sigma^{\rm imp}(T,{\bf c})$ is directly related
to the impurity assisted tunneling. It is convenient to write it down
as a sum of two contributions:
\[
\sigma^{\rm imp}(T,{\bf c}) =
\sigma^{\rm imp}_{\rm el}(T,{\bf c}) +
\sigma^{\rm imp}_{\rm sf}(T,{\bf c}),
\]
where the first term corresponds to the conductivity due to
elastic spin-conserving processes of scattering electron on the
impurity and the second one summarizes all other events when the
conduction electron changes its spin after tunneling through
the barrier.  We have derived the analytical expressions for
these two terms, which are valid under two assumptions. Namely,
under domination of single electron scattering on impurities over
multiple scattering of two and more electrons on the same center
and under absence of polarization of impurity spin induced by
the ejection of spin-polarized electrons. Then the final result for
these terms is written as (the details of its derivation are outlined further)
\begin{eqnarray}
\sigma^{\rm imp}_{\rm el}(T,{\bf c}) =
\frac{1}{S}
\left(\frac{2e^2}{\pi\hbar}\right)
\int_{-\infty}^{+\infty} d\varepsilon
\left\{
-\frac{\partial
f_{\uparrow}(\varepsilon - \mu_B H_z^{\rm eff}) }
{\partial\varepsilon}
\Bigl\langle(\hat{t}^{\uparrow}_z(\varepsilon) )^{\dagger}
\hat{t}^{\uparrow}_z(\varepsilon) \Bigr\rangle
\Phi_{\uparrow}^{L}({\bf c}) \Phi_{\uparrow}^{R}({\bf c}) - \right.
\nonumber
\\
\left.
\frac{\partial
f_{\downarrow}(\varepsilon + \mu_B H_z^{\rm eff}) }
{\partial\varepsilon}
\Bigl\langle( \hat{t}^{\downarrow}_z(\varepsilon) )^{\dagger}
\hat{t}^{\downarrow}_z(\varepsilon)\Bigr\rangle
\Phi_{\downarrow}^{L}({\bf c}) \Phi_{\downarrow}^{R}({\bf c})
\right\},
\label{s_imp}
\\
\sigma^{\rm imp}_{\rm sf}(T,{\bf c}) =
\frac{1}{S}
\left(\frac{2e^2}{\pi\hbar}\right)
\frac{1}{k_B T}
\int_{-\infty}^{+\infty} d\varepsilon
\Bigl\{
f_{\uparrow}(\varepsilon - \mu_B H_z^{\rm eff})
\left[1 - f_{\downarrow}(\varepsilon + \mu_B H_z^{\rm eff})\right]
\Bigl\langle \hat{t}_{-}(\varepsilon)
\hat{t}_{+}(\varepsilon)\Bigr\rangle +
\nonumber
\\
f_{\downarrow}(\varepsilon + \mu_B H_z^{\rm eff})
\left[1 - f_{\uparrow}(\varepsilon - \mu_B H_z^{\rm eff})\right]
\Bigl\langle \hat{t}_{+}(\varepsilon)
\hat{t}_{-}(\varepsilon)\Bigr\rangle
\Bigr\}\times
\frac{1}{2}\left\{
\Phi_{\uparrow}^{L}({\bf c}) \Phi_{\downarrow}^{R}({\bf c}) +
\Phi_{\downarrow}^{L}({\bf c}) \Phi_{\uparrow}^{R}({\bf c})
\right\}.
\nonumber
\end{eqnarray}
Here $S$ is the junction area,
$\Phi_{\uparrow(\downarrow)}^{L}({\bf c})$ and
$\Phi_{\uparrow(\downarrow)}^{R}({\bf c})$ are the probabilities
of tunneling of the electron from the left or from the right electrode
to impurity, located at point ${\bf c}$.
Omitting the exponentially small terms, the expression for these
probabilities can be written as
\begin{eqnarray}
\Phi_{\mu}^{L}({\bf c}) &=&
\int_0^{x_{\rm max}^{\mu}} \frac{xdx}{2\pi}
\frac{2 c_1^{\mu} m_0^2 q_0^2 }
{(m_0^2 c_1^{\mu\,2} + q_2^2)} e^{-2q_2(c-a)},
\nonumber
\\
\Phi_{\rho}^{R}({\bf c}) &=&
\int_0^{x_{\rm max}^{\rho}} \frac{xdx}{2\pi}
\frac{2 c_3^{\rho} m_0^2 q_0^2 }
{(m_0^2 c_3^{\rho\,2} + q_2^2)} e^{-2q_2(b-c)}.
\nonumber
\end{eqnarray}
The quantities $\langle(\hat t^{\uparrow(\downarrow)}_z )^{\dagger}
\hat t^{\uparrow(\downarrow)}_z
\rangle(\varepsilon)$
and $\langle \hat t_{-}\hat t_{+}\rangle(\varepsilon)$,
$\langle \hat t_{+} \hat t_{-}\rangle(\varepsilon)$
in~(\ref{s_imp}) represent
the scattering amplitudes of electron on the impurity center
for the case of spin conserving
$(|in,\uparrow\rangle \to \\ |out,\uparrow\rangle)$ or
$|in,\downarrow\rangle \to |out,\downarrow\rangle)$
and spin-flip
$(|in,\uparrow\rangle \to |out,\downarrow\rangle$ or
$|in,\downarrow\rangle \to |out,\uparrow\rangle) $
transitions averaged over the distribution of paramagnetic
impurity spin. Here $|in\rangle$ and $|out\rangle$ denote
the initial and final states of impurity.
Operators
$\hat t_z^{\uparrow(\downarrow)}$ and
$\hat t_{\pm}$ form a one-center
matrix
$\hat{t} = \left(
{\hat t_z^{\uparrow} \atop \hat t_{+} }
{\hat t_{-} \atop \hat t_z^{\downarrow} } \right) $
in the direct product of
the linear subspaces of electron's and impurity's spins
and are expressed as
\begin{eqnarray}
{} &\displaystyle
\hat t_z^{\uparrow(\downarrow)}(\varepsilon) =
\frac{1}
{1 - \hat V_z^{\uparrow(\downarrow)}(\varepsilon)
G_{\uparrow(\downarrow)}(\varepsilon)}
\hat V_z^{\uparrow(\downarrow)}(\varepsilon),& {}
\nonumber
\\
{} &\displaystyle
\hat t_{\pm}(\varepsilon) = -
\frac{1}
{1 - \hat V_z^{\downarrow(\uparrow)}(\varepsilon)
G_{\downarrow(\uparrow)}(\varepsilon)}
\hat S_{\pm} \frac{a_0^3 J/2}
{1 - a_0^3 (\varepsilon_0 + \frac{1}{2} J \hat S_z)
G_{\uparrow(\downarrow)}(\varepsilon)},& {}
\label{t-matrix}
\end{eqnarray}
where effective potentials
$\hat V_z^{\uparrow(\downarrow)}$ are given by
\[
\hat V_z^{\uparrow(\downarrow)}(\varepsilon) = a_0^3\left\{
\varepsilon_0 \mp \frac{1}{2} J \hat S_z + \frac{1}{4} \hat S_{\mp}
\frac{ a_0^3 G_{\downarrow(\uparrow)}(\varepsilon) J^2}
{1 - a_0^3 (\varepsilon_0 \pm \frac{1}{2} J \hat S_z)
G_{\downarrow(\uparrow)}(\varepsilon)} \hat S_{\pm} \right\}.
\]
Here $G_{\uparrow(\downarrow)}(\varepsilon)$
is the electron Green function at point ${\bf c}$:
$$
G_{\mu}(\varepsilon,{\bf c}) =
\int_{0}^{\kappa_{\rm max}}
G_{\mu\kappa}(\varepsilon,{\bf c})\,
\frac{\kappa d\kappa}{2\pi},
$$
where $\kappa_{\rm max} = \frac{ 2\sqrt{\pi} }{a_0} $ is a cut-off of
in plane momentum that stems from the finite size of Brillouin zone
(we substitute the Brillouin zone's projection onto
$(\kappa_x, \kappa_y)$ plane by the circle of radius
$\kappa_{\max}$ of the same square in $\kappa_{\parallel}$--plane).
The real and imaginary part of
$G_{\kappa}^{\mu}(\varepsilon,{\bf c})$ ($\mu$ is the spin index) in the
leading order of magnitude are given by
\begin{eqnarray}
{} &\displaystyle
{\rm Re}\,G_{\kappa}^{\mu}(\varepsilon,{\bf c}) = -
\frac{m_0}{q_2}, & {}
\label{G_c}
\\
{} &
{\rm Im}\,G_{\kappa}^{\mu}(\varepsilon,{\bf c}) = -
\left(
\Phi_{\mu}^{L}({\bf c}) + \Phi_{\mu}^{R}({\bf c})
\right). & {}
\nonumber
\end{eqnarray}

\begin{figure}
\begin{center}
\includegraphics[scale=0.7, angle=0]{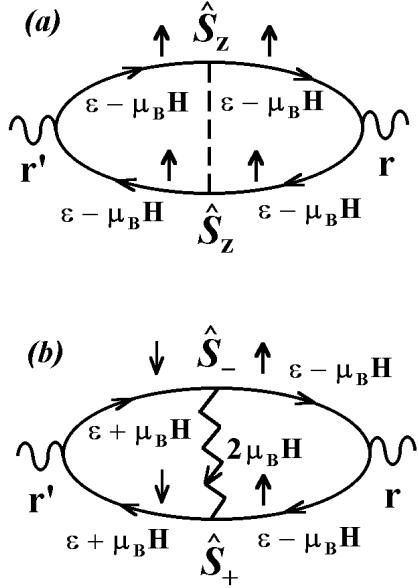}
\end{center}
\caption{\small Two diagrams that make contribution to the conductivity
in the second order of perturbation theory in case of $a)$ elastic spin-
conserving scattering and 
$b)$ non-elastic spin-flip scattering.}
\end{figure}

  Let us now explain the derivation of expression~(\ref{s_imp})
and clarify  the two assumptions under which this formula
is valid. To derive~(\ref{s_imp})
from the starting point~(\ref{Kubof}) one can
first of all examine two diagrams (a) and (b)
(see Fig.~2) that contribute to spin-conserving and spin-flip
part of $\sigma^{\rm imp}(T,{\bf c})$ at second order of $J$,
respectively. One may easy verify that a general structure
of these diagrams is just the same as the final
result in form~(\ref{s_imp}) with the
mere difference that the one-center $t$--matrix is reduced
at first order of $J$ to the initial potential
$\varepsilon_0 - \frac{J}{2}
\left( {\hat S_z \atop \hat S_{+} }{\hat S_{-} \atop -\hat S_z} \right)$.

 Moreover, the diagram (b) contains both direct and indirect
processes in equal proportion with common factor 1/2 for any of the
possible channels
$|in,\uparrow\rangle \to |out,\downarrow\rangle$ or
$|in,\downarrow\rangle \to |out,\uparrow\rangle$.
The thermodynamic  averaging $\langle\dots\rangle$ in the second-order
expansion is simply reduced to the averaging over the Boltzman
distribution of the impurity spin in the "external" effective
magnetic field $H_z^{\rm eff}$ which was introduced
in (\ref{Hamilt}), i.e. with the density matrix
$\hat \rho_0 = Z^{-1}\exp
\left\{\frac{2\mu_B H_z^{\rm eff} \hat S_z}{k_B T}\right\}$,
where
$Z = 2\cosh \left(\frac{2\mu_B H_z^{\rm eff}}{k_B T}\right) + 1$.
After that, it is easy to check that the total probabilities
(with account of Fermi factors of electron states)
of direct and inverse processes are equal
which means that the principle of detailed equilibrium holds.
In particular it leads to the vanishing of spin-flip processes
in a system at zero temperature and vanishing voltage
bias.

  After that preliminary discussion two assumptions should be
made to justify the result~(\ref{s_imp}):

 i) We assume that the occupation
of the given impurity center simultaneously by two electrons
with different spin (due to Pauli principle) is a rather
rare event or, in other words, we do not take into account
many-electrons effects. It may be justified:
a) by Coulomb interaction between electrons that make unprofitable
their arrangement at the same site of the lattice;
b) by the large number of impurity centers that provides a sufficiently
large number of one-step channels so that electrons may be
considered as independent.

  ii) We also neglect the influence of electron current on the
statistical distribution of paramagnetic spins inside the oxide barrier.
This assumption is valid for practical intensity of tunneling current
which is low enough not to produce a spin polarization
of impurities by injection of spin-polarized charge carriers.

  Under these assumptions the expression~(\ref{s_imp}) can be obtained by
simple substitution of scattering potential $\hat H_{\rm int}$ at site
${\bf c}_i$ on Fig.~2 by the corresponding one-center $t$--matrix
in accordance with~(\ref{t-matrix})
and assuming that the averaging over the degrees of freedoms
of the impurity is carried out by means of unperturbed
density matrix
$\hat \rho_0 = Z^{-1} \exp\left\{ \frac{2\mu_B H_z^{\rm eff}
\hat S_z}{k_B T}\right\}$.
In this form the structure of the result~(\ref{s_imp})
is similar to the one obtained in Ref.~\cite{Levy1997},
where the spin-flip scattering of electrons
at interfaces of tunnel junctions was investigated in the framework
of tunneling Hamiltonian and the second order
perturbation theory.

\subsection{Resonant tunneling in the case of nonmagnetic impurities}
  To extract the physical nature of resonant tunneling through the impurity
states contained in expression~(\ref{s_imp})
we proceed as follows. For the sake
of clarity and simplicity we consider first the case of zero-spin
impurity. Then only one element of $t$--matrix at cite ${\bf c}$
survives
\[
t_0^{\uparrow(\downarrow)}(\varepsilon) =
\frac{a_0^3\,\varepsilon_0}{1 - a_0^3 \varepsilon_0
G_{\uparrow(\downarrow)}(\varepsilon)}.
\]
It defines the position of a resonant level inside the gap
of the dielectric band structure by finding the root of the equation
$a_0^3\varepsilon_0 {\rm Re}G_{\uparrow(\downarrow)}(\varepsilon_i) = 1$.
From expression~(\ref{G_c}), it follows that the real part
of the Green function ${\rm Re}\,G_{\mu}(\varepsilon,{\bf c})$ is
independent on {\bf c} and spin $\mu$ up to exponentially
small terms. Therefore, the position of level $\varepsilon_i$
is weakly depend both on the position of impurity inside the
barrier and on the direction of the spin of tunneling electron.
Evidently, only those impurities for which $\varepsilon_i$
is close to the chemical potential $\varepsilon_F$ contribute
to a significant extent of the total current at low bias voltage.
Therefore, it is possible to expand
the denominator in $t_0^{\uparrow(\downarrow)}(\varepsilon)$
in powers of $(\varepsilon_i - \varepsilon)$. If we now
introduce the position dependent line-widths
\begin{equation}
\Gamma_{\mu}^L({\bf c}) = \Phi_{\mu}^L({\bf c})/ {\rm Re}\, G'(\varepsilon_F),
\quad
\Gamma_{\mu}^R({\bf c}) = \Phi_{\mu}^R({\bf c})/ {\rm Re}\, G'(\varepsilon_F),
\label{lines}
\end{equation}
where ${\rm Re}\, G'(\varepsilon_F) =
\frac{\partial}{\partial\varepsilon} {\rm Re}\, G(\varepsilon)
\Bigl|_{\varepsilon = \varepsilon_F} $
is the energy derivative of the electron Green function
at Fermi level, then
we obtain the general formula for the resonant case of impurity
assisted tunneling
\begin{equation}
\sigma^{\rm imp}({\bf c}) \simeq \frac{1}{S}
\left(\frac{2 e^2}{\pi\hbar}\right)
\sum_{\mu}
\int_{-\infty}^{+\infty}
\sum_i
\frac{\Gamma_{\mu}^L({\bf c})\Gamma_{\mu}^R({\bf c})}
{(\varepsilon_F - \varepsilon_i)^2 + \Gamma_{\mu}^{2}({\bf c})}
\left(-\frac{\partial f(\varepsilon)}{\partial \varepsilon}\right)\,
d\varepsilon,
\label{s_res}
\end{equation}
where
$\Gamma_{\mu}({\bf c}) =
\Gamma_{\mu}^L({\bf c}) + \Gamma_{\mu}^R({\bf c})$  and
the summation by $i$ is performed
over all resonant levels. For the qualitative
analysis, one may evaluate the expressions~(\ref{lines}) for
$\Gamma^{R(L)}_{\mu}({\bf c})$ approximately by considering
the case $\kappa = 0$ which is valid
if $e^{-2q_0w}\ll 1$. In this approximation
\begin{eqnarray}
\Gamma^L_{\mu}({\bf c} ) & = &
\frac{2k^F_{1\mu}m_0}{m_0^2 k^{F\,2}_{1\mu} + q_0^2}
\left(\frac{q_0^2}{2m_0}\right)
\frac{e^{-2q_0(c-a)}}{c-a},
\nonumber
\\
\Gamma^R_{\mu}({\bf c}) & = &
\frac{2k^F_{3\mu}m_0}{m_0^2 k^{F\,2}_{3\mu} + q_0^2}
\left(\frac{q_0^2}{2m_0}\right)
\frac{e^{-2q_0(b-c)}}{b-c},
\label{Gamma}
\end{eqnarray}
and expression~(\ref{s_res}) reproduces the result of Ref.~\cite{Larkin}.
To proceed further, we discuss some assumptions
concerning the parameters of the model. We consider the case
of Co electrode and Al$_2$O$_3$ as an oxide layer
and take typical values of $k^F_{\uparrow} = 1.09$\,\AA$^{-1}$,
$k^F_{\downarrow} = 0.42$\,\AA$^{-1}$, $m\approx 1$ for itinerant
electrons in Co and a typical barrier height
for Al$_2$O$_3$ (measured from the Fermi level $\varepsilon_F$)
$U_0 - \varepsilon_F = 3\,{\rm eV}$ with an effective mass
$m_0 = 0.4$ (Ref.~\cite{Bratkovsky}),
that gives $q_0 \simeq 0.56$\,\AA$^{-1}$.
Assuming the thickness of the barrier $w \simeq 20$\,\AA$^{-1}$,
one may estimate the conductance $\sigma_0$ of the system
without impurity from~(\ref{sigma0}) by means of
approximate formula
\begin{equation}
\sigma_0 \simeq \frac{2e^2}{\pi^2\hbar}\left(\frac{q_0}{w}\right)
\sum_{\mu}
\frac{k_{1\mu}^F k_{3\mu}^F q_0^2 m_0^2 e^{-2q_0 w} }
{(m_0^2 k_{1\mu}^{F 2} + q_0^{2})(m_0^2 k_{3\mu}^{F 2} + q_0^{2})},
\label{s0_approx}
\end{equation}
that leads to GMR~$\simeq$\,16\,\%. To estimate the value of
the line-width~(\ref{Gamma}) we consider impurities located close
to left interface at a distance, say, of two atomic layers which
corresponds to $(c - a)\simeq 4$\,\AA. For spin up electrons
it gives $\Gamma_{\uparrow}({\bf c})\simeq 0.02$\,eV.
Further in this paper we restrict ourselves
to the case of temperature interval from $4.2 - 300$~K
(0.025~eV).
We assume that the impurity levels $\varepsilon_i$
in the band gap of the insulator form
a narrow impurity band of width $\Delta\varepsilon$
which spreads symmetrically with respect to Fermi level
$\varepsilon_F$ and, following Ref.~\cite{Larkin}, we introduce
its density of states $\nu(\varepsilon)$ per unit volume and
unit energy interval. We assume that $\Delta\varepsilon$
is of the order $0.1$ to $0.2$~eV, i.e.
an order of magnitude greater that the above estimated line-width.
In this context, with a good accuracy, the impurity
conductance~(\ref{s_res}) rewrites as follows:
\begin{equation}
\sigma^{\rm imp}({\bf c}) \simeq
\frac{2e^2}{\hbar}\nu(\varepsilon_F)
\sum_{\mu}
\int_{-\infty}^{+\infty}
\left( -\frac{\partial f}{\partial \varepsilon}\right)
\frac{ \Gamma_{\mu}^L({\bf c})\Gamma_{\mu}^R({\bf c})}
{\Gamma_{\mu}({\bf c})}\rho(\varepsilon,{\bf c})
\,d\varepsilon,
\label{s_Larkin}
\end{equation}
where factor
\[
\rho(\varepsilon,{\bf c}) =
\frac{2}{\pi}\arctan\left(\frac{\Delta\varepsilon}{2\Gamma_{\mu}({\bf 
c})}\right)
\]
arises from the integration of exp.~(\ref{s_res}) over impurity
levels $\varepsilon_i$ in the range of impurity band.
Due to above mentioned estimations equality
$\rho(\varepsilon, {\bf c})\simeq 1$ holds with a good degree
of accuracy and in this case exp.~(\ref{s_Larkin}) becomes
in agreement with Refs.~\cite{Larkin,Bratkovsky}.

\subsection{Resonant tunneling in the case of paramagnetic impurities}

 To investigate the general case of paramagnetic impurity we follow
the same procedure as in the previous section. Let
${\bf J} = {\bf s} + {\bf S}$ be the total magnetic moment
of the system. We may state that $[H,J_z] = 0$ and, therefore,
$J_z$ is a good quantum number. We regard the
$\hat{t}$--matrix~(\ref{t-matrix})
as an operator acting on the spinor subspace
$|\sigma,m \rangle$, where $\sigma = \pm\frac{1}{2}$ and
$m = \pm 1, 0$ corresponds to the projection of the $z$--component
of electron and impurity spin, respectively
(we consider the case ${\bf S} = 1$). As far as its total magnetic
moment along the $z$--axis $J_z = s_z + S_z$ is conserved,
the matrix elements
$\langle\sigma_1 m_1|\hat{t}|\sigma_2 m_2 \rangle$ are non-zero
only if $m_1 + \sigma_1 = m_2 + \sigma_2$ and, therefore,
it is convenient to introduce the notation
$t_{m_j}^{\sigma_1\sigma_2} =
\langle\sigma_1 m_1|\hat{t}|\sigma_2 m_2 \rangle $,
where
$ m_j = m_1 + \sigma_1 = m_2 + \sigma_2$.
These elements are simply calculated
from~(\ref{t-matrix}). The non-zero
ones are written as follows
\begin{eqnarray}
t^{\uparrow\uparrow}_{3/2} & = &
\frac{a_0^3(\varepsilon_0 - J/2)}
{1 - a_0^3 (\varepsilon_0 - J/2)G_{\uparrow}(\varepsilon)},
\nonumber
\\
t^{\downarrow\downarrow}_{-3/2} & = &
\frac{a_0^3(\varepsilon_0 - J/2)}
{1 - a_0^3 (\varepsilon_0 - J/2)G_{\downarrow}(\varepsilon)},
\label{t_z}
\end{eqnarray}
and
$
\hat{t}_{\pm 1/2} =
\left( { t^{\downarrow\downarrow}_{\pm 1/2}\atop
 t^{\uparrow\downarrow}_{\pm 1/2} }
{ t^{\downarrow\uparrow}_{\pm 1/2} \atop
 t^{\uparrow\uparrow}_{\pm 1/2} }
\right)
$
corresponding to the subspace $m_j = \pm \frac{1}{2}$ with
\begin{eqnarray}
t^{\downarrow\downarrow}_{1/2}
(t^{\uparrow\uparrow}_{-1/2}) & = &
\frac{a_0^3}{\Delta_{\pm 1/2}(\varepsilon)}
\left\{ \varepsilon_0 + J/2 -
a_0^3 G_{\uparrow(\downarrow)}(\varepsilon)(\varepsilon_0 - J/2)
(\varepsilon_0 + J)
\right\},
\nonumber
\\
t^{\uparrow\uparrow}_{1/2}
(t^{\downarrow\downarrow}_{-1/2}) & = &
\frac{a_0^3}{\Delta_{\pm 1/2}(\varepsilon)}
\left\{ \varepsilon_0  -
a_0^3 G_{\downarrow(\uparrow)}(\varepsilon)(\varepsilon_0 - J/2)
(\varepsilon_0 + J)
\right\},
\label{t_12}
\\
t^{\uparrow\downarrow}_{\pm 1/2} & = &
t^{\downarrow\uparrow}_{\pm 1/2} =
- \frac{a_0^3}{\sqrt{2}\Delta_{\pm 1/2}}J,
\nonumber
\end{eqnarray}
where denominators are
\[
\Delta_{\pm 1/2}(\varepsilon) =
\Bigl(1 - a_0^3 G_{\downarrow}(\varepsilon)(\varepsilon_0 - J/2) \Bigr)
\Bigl(1 - a_0^3 G_{\uparrow}(\varepsilon)(\varepsilon_0 + J) \Bigr) \pm
a_0^3 J (G_{\uparrow}(\varepsilon) - G_{\downarrow}(\varepsilon) ).
\]
As can be seen, two poles of the $\hat{t}$--matrix defined from
equations
$a_0^3 {\rm Re}\,G(\varepsilon_{3/2})(\varepsilon_0 - J/2) = 1$
and
$a_0^3 {\rm Re}\,G(\varepsilon_{1/2})(\varepsilon_0 + J) = 1$ correspond
to two multiplets $\varepsilon_{3/2}$  and $\varepsilon_{1/2}$
with a total angular momentum $j = 3/2$ and $j = 1/2$,
respectively. If $J>0$, then
$\varepsilon_{3/2} < \varepsilon_{1/2}$, i.e. the multiplet with
$j = 3/2$ has a lower energy than one with $j = 1/2$.
As for non-magnetic impurity, we restrict ourselves
by considering the regime of only one-channel resonant tunneling.
We assume that $J>0$ and the lowest impurity levels
$\varepsilon_{i} = \varepsilon_{3/2}$ corresponding to the
multiplet with $j = 3/2$ lie close to $\varepsilon_F$.
We note that the typical value of exchange coupling $J$ is of order 1~eV
and due to this fact we may eliminate the resonant level $\varepsilon_{1/2}$
from further consideration. Then, as in the previous analysis for a non-
magnetic
impurity, only the resonant part of the $\hat{t}$--
matrix~(\ref{t_z},\ref{t_12})
at energies close to chosen $\varepsilon_{3/2}$
is essential for the subsequent calculations. Expressions~(\ref{t_z},\ref{t_12}) 
can be easily written as follows:
\begin{eqnarray}
{} &\displaystyle
t^{\uparrow\uparrow(\downarrow\downarrow)}_{\pm 3/2}(\varepsilon) =
\frac{1}{G'(\varepsilon)}
\frac{1}{\varepsilon - \varepsilon_i +
i\Gamma_{\uparrow(\downarrow)}({\bf c})}, & {}
\nonumber
\\ {} & { \displaystyle
\hat{t}_{1/2}(\varepsilon) = \frac{1}{G'(\varepsilon)}
\frac{1}{\varepsilon - \varepsilon_i +
i\gamma_{\uparrow}({\bf c})}
\left(
\begin{array}{cc}
\frac{1}{3} & \frac{\sqrt{2}}{3} \\
\frac{\sqrt{2}}{3}  & \frac{2}{3}
\end{array}
\right), } & {}
\label{t_res}
\\
{} & {  \displaystyle
\hat{t}_{-1/2}(\varepsilon) =  \frac{1}{G'(\varepsilon)}
\frac{1}{\varepsilon - \varepsilon_i +
i\gamma_{\downarrow}({\bf c})}
\left(
\begin{array}{cc}
\frac{2}{3} & \frac{\sqrt{2}}{3} \\
\frac{\sqrt{2}}{3} & \frac{1}{3}
\end{array}
\right), } & {}
\nonumber
\end{eqnarray}
where
$\gamma_{\uparrow}({\bf c}) = \frac{2}{3}
\Gamma_{\uparrow}({\bf c}) + \frac{1}{3}\Gamma_{\downarrow}({\bf c})$,
$\gamma_{\downarrow}({\bf c}) = \frac{1}{3}
\Gamma_{\uparrow}({\bf c}) + \frac{2}{3}\Gamma_{\downarrow}({\bf c})$
are the inverse lifetimes of the resonant states with
$m_j = \pm 1/2$. This result allows simple qualitative
interpretation. Let us look, for a example, at quantum states
with $m_j = 1/2$. From elementary quantum mechanical theory
one may conclude that
\begin{eqnarray}
{} &
\phi^{\uparrow}_{1/2} =
\Bigl| \uparrow, m_s = 0 \Bigr\rangle =
\sqrt{\frac{2}{3}} \Bigl|j= \frac{3}{2},
m_j = \frac{1}{2}\Bigr\rangle +
\sqrt{\frac{1}{3}} \Bigl|j=\frac{1}{2},
m_j = \frac{1}{2}\Bigr\rangle, & {}
\nonumber
\\
{} &
\phi^{\downarrow}_{1/2} =
\Bigl| \downarrow, m_s = 1 \Bigr\rangle =
\sqrt{\frac{1}{3}} \Bigr |j=\frac{3}{2},
m_j = \frac{1}{2} \Bigl \rangle -
\sqrt{\frac{2}{3}} \Bigr |j=\frac{1}{2},
m_j = \frac{1}{2} \Bigl \rangle. & {}
\label{phi}
\end{eqnarray}
As we have assumed, only
$\Bigr |j=\frac{3}{2}, m_j = \frac{1}{2} \Bigl \rangle$
gets into resonance and, therefore, e.g.
$t^{\uparrow\downarrow}_{1/2} \sim
\Bigl\langle \phi^{\uparrow}_{1/2}\Bigr|\,\hat{t}\,\Bigl|
\phi^{\downarrow}_{1/2}\Bigr\rangle \sim \sqrt{2}/3
$
in agreement with~(\ref{t_res}). On the other hand,
\par
\begin{center}
$
\Bigr |j=\frac{3}{2}, m_j = \frac{1}{2} \Bigl \rangle =
\sqrt{\frac{2}{3}}\Bigr|\uparrow, m_s = 0 \Bigl \rangle +
\sqrt{\frac{1}{3}}\Bigr|\downarrow, m_s = 1 \Bigl \rangle
$
\end{center}
\par
\noindent
and, hence, its inverse life time is given by
$\tau^{-1}_{1/2} = \frac{2}{3}\tau^{-1}_{\uparrow} +
\frac{1}{3}\tau^{-1}_{\downarrow}$.

 We substitute all $\hat{t}$--matrix
elements in~(\ref{s_imp}) by its
resonance expansion~(\ref{t_res}). To proceed further, one has to perform
in~(\ref{s_imp}) the configuration averaging over all impurity
centers and thermodynamic one over all possible channels.
Suppose, that the impurities are distributed uniformly
in the space in the interval
$[z_0 - \Delta, z_0 + \Delta]$ along $z$--direction with the
width of $2\Delta$ and center $z_0$ which we have chosen to be close
to left (L) ferromagnetic contact.
After that, it is possible, first of all, to average the
Lorentzian peaks over the distribution of impurity levels
by averaging them over $\varepsilon_i$ in the range of impurity band
with factor $\nu(\varepsilon_F)$ and to perform the thermodynamic
averaging by integrating over $\varepsilon$ and neglecting
the dependencies of $\Gamma^{L(R)}({\bf c})$ on energy. On the third
step, the averaging over the space distribution of impurities
along $z$--direction should be made. Following the outlined
procedure, the total conductance~(\ref{s_imp})
as a function of temperature
for the parallel and antiparallel alignment of magnetization
of the ferromagnetic layers is written as a sum of factorized
terms over all possible scattering channels:
\begin{eqnarray}
{} & \sigma^{P}(T) = {\displaystyle \frac{2e^2}{\hbar} }
{\displaystyle \sum_{\mu\rho\,m_j}}
P^{\mu\rho}_{m_j}\left(\frac{\mu_B H_z^{\rm eff}}{k_B T}\right)
{^{P}\sigma_{m_j}^{\mu\rho}}(z_0,\Delta)\nu(\varepsilon_F) +\
\sigma_0^P, & {}
\nonumber
\\
{} & \sigma^{AP}(T) = {\displaystyle \frac{2e^2}{\hbar} }
{\displaystyle \sum_{\mu\rho\,m_j} }
P^{\mu\rho}_{m_j}\left(\frac{\mu_B H_z^{\rm eff}}{k_B T}\right)
{^{AP}\sigma_{m_j}^{\mu\rho}}(z_0,\Delta)\nu(\varepsilon_F) +\
\sigma_0^{AP},& {}
\label{s_T}
\end{eqnarray}
where
\[
\sigma^{\mu\rho}_{m_j}(z_0,\Delta) = \frac{1}{2\Delta}\int_{z_0-
\Delta}^{z_0+\Delta}
\sigma^{\mu\rho}_{m_j}(c)\rho(\varepsilon_F,c) \,dc.
\]
The origine of $\rho(\varepsilon_F,c)$ is the same
as in exp.~(\ref{s_Larkin}) and
explicit form of functions $P^{\mu\rho}_{m_j}(h)$ and
$\sigma^{\mu\rho}_{m_j}(c)$ is given in Appendix A. We have
also used the same notation of matrix indexes as it was
previously introduced for $\hat{t}$--matrix elements.
$\sigma_0^{P}$ and $\sigma_0^{AP}$ are the tunnel conductances
of the pure system in accordance with~(\ref{sigma0}).
Factors $P_{m_j}^{\mu\rho}$ and $\sigma_{m_j}^{\mu\rho}$
represent the thermodynamic and quantum mechanical probabilities
of the given process, respectively. Expressions (\ref{s_T})
are the final results of this section and their analysis is
presented below (see Sec.~III).

\subsection{Dependence of conductivity on bias-voltage}

  We are also interested in $I(V)$ characteristic of
the considered system. To derive the general formula for the current, one may 
simply extend the expressions~(\ref{s_imp}) to the case of finite applied
bias voltage. Consider, for example, the contribution
to the total current $I$, coming from all possible channels of the form
$|in, \uparrow \rangle \to |out, \downarrow \rangle$ for
tunnel electrons moving from the left electrode
to the right electrode and of the form
$|in, \downarrow \rangle \to |out, \uparrow \rangle$
for electrons moving from the right to the left,
respectively, i.e. in both cases an electron has an "up" projection
of spin in the left contact and "down" projection of spin
in the right one after or before scattering.
From the general concept, one may conclude
that this contribution to the current can be written as
\begin{eqnarray}
I^{\uparrow\downarrow}(V) =
\frac{1}{S}
\left(\frac{2e^2}{\pi\hbar}\right)\int_{-\infty}^{+\infty} d\varepsilon
\biggl\{
f_{\uparrow}(\varepsilon - \mu_B H_z^{\rm eff} - eV)
\left[1 - f_{\downarrow}(\varepsilon + \mu_B H_z^{\rm eff})\right]
\Bigl\langle \hat t_{-}(\varepsilon)
\hat t_{+}(\varepsilon)\Bigr\rangle -
\nonumber
\\
f_{\downarrow}(\varepsilon + \mu_B H_z^{\rm eff})
\left[1 - f_{\uparrow}(\varepsilon - \mu_B H_z^{\rm eff} - eV)\right]
\Bigl\langle \hat t_{+}(\varepsilon)
\hat t_{-}(\varepsilon)\Bigr\rangle
\biggr\}
\Phi_{\uparrow}^{L}({\bf c}) \Phi_{\downarrow}^{R}({\bf c}),
\label{I_V}
\end{eqnarray}
where it is assumed that the voltage bias is applied from the
left to the right direction. It is important to notice
that inelastic spin-flip processes of the electron scattering
on the impurity were taken into account in derivation of the
exp.~(\ref{I_V}) but they were omitted in Ref.~\cite{Guinea}.
Analogous expressions can be written for
all other channels. In the case under consideration, expression
(\ref{I_V}) contains two regimes of non-linear behavior of
$I(V)$ characteristic. The first one reproduces a zero bias
anomaly due to excitation of spin-flip processes at low bias
voltages of order of magnitude $\mu_B H^{\rm eff}_z$
(we believe that it is of order 5~mV). In this range,
as before, one may assume that the resonance amplitudes
$\langle
(\hat t_z^{\uparrow(\downarrow)})^{\dagger}
(\hat t_z^{\uparrow(\downarrow)}) \rangle $
and
$\langle \hat t_{-}\hat t_{+}\rangle$ are nearly independent
of the energy after averaging over all possible configurations
of impurities. As a result, the voltage dependence
of total currents for parallel and antiparallel configurations
are given by formulae similar to~(\ref{s_T}):
\begin{eqnarray}
I^{P}(V,T) & = &
\frac{2e}{\hbar}\sum_{\mu\rho\,m_j}
I^{\mu\rho}_{m_j}(V, H_z^{\rm eff})\,
{^{P}\sigma_{m_j}^{\mu\rho}}(z_0,\Delta)\nu(\varepsilon_F) +\
\sigma_0^P V,
\nonumber
\\
I^{AP}(V,T) & = &
\frac{2e}{\hbar}\sum_{\mu\rho\,m_j}
I^{\mu\rho}_{m_j}(V, H_z^{\rm eff})\,
{^{AP}\sigma_{m_j}^{\mu\rho}}(z_0,\Delta)\nu(\varepsilon_F) +\
\sigma_0^{AP} V.
\label{I_VT}
\end{eqnarray}
The expressions for $I^{\mu\rho}_{m_j}(V, H_z^{\rm eff})$
are given in Appendix B.
The voltage dependent conductances $\sigma^P(V,T)$ and
$\sigma^{AP}(V,T)$ can be obtained from~(\ref{I_V}) by derivation
with respect to $V$. The detailed analysis of this physical
situation is presented in the next section.

  The second source of possible non-linear character of $I(V)$
dependence is the variation of potential profile $U(z)$
(see Fig.~1) under applied bias voltage. It follows, then,
that the latter introduces the correction to exp.~(\ref{Gamma})
and (\ref{s0_approx}) and they can be calculated with the use
of Wentzel--Kramers--Brillouin (WKB) approximation~\cite{WKB} assuming
that the applied voltage produces the uniform electrical
field inside the insulating layer. In the case of pure
tunnel conductance it is known\cite{Vedy2000} that
both conductances for the parallel and antiparallel
configurations increase with the increase of applied
voltage so that the TMR as a function of $V$, defined
as $\displaystyle\frac{I^P(V) - I^{AP}(V)}{I^{AP}(V)}$,
drops significantly at the voltages of order 1~eV.
The contribution of impurity assisted tunneling
may change considerably this situation in the case
of non-uniform spatial distribution of impurities,
e.g.~when they are distributed in the vicinity
of only one electrode. In this particular
situation, as we will show, the essential
variation of TMR amplitude in the case of magnetic
impurities (in contrast to non-magnetic ones)
will take place at the bias voltages compared
with impurity band width $\Delta\varepsilon$.

 For the sake of simplicity we consider, first, the case of
non-magnetic impurities. In the WKB approximation the contribution
from all impurities, located at given point ${\bf c}$, to the
total current $I(V)$ will have the form similar to
(\ref{Gamma}) and (\ref{s_Larkin}):
\begin{equation}
j^{\rm imp}({\bf c}) = \frac{2e^2}{\hbar}\nu(\varepsilon_F)
\sum_{\mu}\int_{-\infty}^{+\infty}
\Bigl\{ f(\varepsilon - eV) - f(\varepsilon)\Bigr\}
\frac{\Gamma^L_{\mu}({\bf c})\Gamma^R_{\mu}({\bf c})}
{\Gamma^L_{\mu}({\bf c}) + \Gamma^R_{\mu}({\bf c})}
\rho(\varepsilon,V)\,d\varepsilon,
\label{j_imp}
\end{equation}
where
\[
\Gamma^L_{\mu}({\bf c}) =
\frac{k^F_{1\mu} q_a m_0 \tau^{-1}_a}
{(q_a^{-})^2 + k_{1\mu}^{F2} m_0^2}e^{-S_a/{\hbar}};
\]
\begin{equation}
\Gamma^R_{\mu}({\bf c}) =
\frac{k^F_{3\mu} q_b m_0 \tau^{-1}_b}
{(q_b^{+})^2 + k_{3\mu}^{F2} m_0^2}e^{-S_b/{\hbar}}\label{G_WKB}.
\end{equation}
Here $q_b^2 = q_0^2 = 2m_0(U-\varepsilon)$,
$q_a^2 = q_0^2 + 2m_0 eV$ are imaginary momenta
of electron with the energy $\varepsilon$
in the vicinity of the right and the left electrode,
$\displaystyle
q_{a(b)}^{\pm} = q_0 \pm \frac12\frac{eEm_0}{q_{a(b)}^2}$,
$E$ is the electric field in the device.
We also introduce
$\displaystyle q_c = q_0 + 2m_0 eV(b-c)/w$
imaginary momenta of electron on the impurity center.
Then $\displaystyle
S_a = \frac{q_a^3 - q_c^3}{3m_0eE}$,
$\displaystyle S_b = \frac{q_c^3 - q_b^3}{3m_0eE}$
represent the classical actions along the path from
the left contact to the point ${\bf c}$ in the barrier
and, afterwards, from this point to the right contact,
respectively;
$\displaystyle\tau_a = \frac{q_a - q_c}{eE}$ and
$\displaystyle\tau_b = \frac{q_c - q_b}{eE}$ denote
the passage times associated with these paths. Factor
\[
\rho(\varepsilon,V) =
\frac{1}{\pi}\left\{
\arctan\left[\frac{\varepsilon - \varepsilon_F - eV\left(\frac{b-c}{w}\right)
+ \frac{\Delta\varepsilon}{2}}{\Gamma^L_{\mu}({\bf c}) + \Gamma^R_{\mu}({\bf 
c})}
\right] -
\arctan\left[\frac{\varepsilon - \varepsilon_F - eV\left(\frac{b-c}{w}\right)
- \frac{\Delta\varepsilon}{2}}{\Gamma^L_{\mu}({\bf c}) + \Gamma^R_{\mu}({\bf 
c})}
\right]
\right\},
\]
as before, arises from the summation over all impurity
levels $\varepsilon_i$ and gives the relative weight
of all resonant channels with energy $\varepsilon$.
To clarify the situation, it is sufficiently to consider
the most resonant channel with energy
$\varepsilon_r = \varepsilon_F + eV(b-c)/w$ at which
$\rho(\varepsilon,V)$ reaches its maximum. One may note that
$\varepsilon_r$ corresponds to the resonant impurity level that exactly 
coincides with Fermi energy at vanishing voltage and it shifts linearly with the
increase of applied bias depending on the position ${\bf c}$ of impurity inside
the barrier. As it was stated earlier, the most interesting case
takes place when the point ${\bf c}$ is situated
close to the left contact. Then one can see
that $\Gamma^L({\bf c})\gg \Gamma^R({\bf c})$
and, thus, $j^{\rm imp}({\bf c}) \sim
\Gamma^R({\bf c})\rho(\varepsilon_r,V)$.
At bias voltages much more lower than the height of the
barrier $\varphi = (U - \varepsilon_F)$, $S_b$ can be expanded
in powers of $V$:
\[
S_b = q_0(b-c)\left\{
1 - \frac{m_0eV}{2q_0^2}\left(\frac{b-c}{w}\right) +
\dots \right\}
\]
which shows that $\Gamma^R({\bf c})\sim\exp(-S_b/{\hbar})$ is
an increasing function of $V$ in the vicinity
of $V=0$. Hence, it leads to increase of differential
conductivity $\sigma(V) = \partial I/\partial V$
under direct bias voltage, applied to the barrier from the left
to the right direction, and to decrease of $\sigma(V)$ under inverse
bias voltage. The physical meaning of such behavior is rather
obvious. From expression for $S_b$ it follows that electrons
tunneling under forward bias due to resonant levels lying close to
$\varepsilon_r$ will propagate through the potential barrier
with the height which is effectively less compared with that
in case of inverse bias.

  The expression for the paramagnetic impurity assisted
current at finite voltages has the structure similar to
exp.~(\ref{j_imp}) with Fermi distribution factors  written
in accordance with the general formula~(\ref{I_V}) and integrand expression
has the form given in Appendix~A, where line-widths 
$\Gamma_{\uparrow(\downarrow)}$ 
have to be substituted by WKB approximation~(\ref{G_WKB}).
In the case of magnetic impurities the above outlined mechanism
of asymmetry in $I(V)$ characteristic due to the resonance levels
will essentially contribute to the voltage dependence of TMR in
question, and it will be discussed in the next section.

\section{Results and discussion}

  In this section we consider the temperature and bias voltage dependencies
of the conductances and TMR effect of the considered structures.
We investigate the case of Co/Al$_2$O$_3$/Co junction with the typical
parameters that have been already mentioned in section 2:
$k^F_{\uparrow}=1.09$\,\AA$^{-1}$,
$k^F_{\downarrow}=0.42$\,\AA$^{-1}$ are the
Fermi momenta of itinerant electrons in
Co, $q_0=0.56$\,\AA$^{-1}$ is the imaginary momentum in the barrier,
$m_0=0.4$ is the effective mass in the insulator and $w=20$\,\AA\ is
its thickness. We focus on the most interesting situation when impurities are 
introduced at the vicinity of the
left electrode at a depth $w_1$ inside the insulator layer. We chose
width $w_1=4.06$\,\AA\ that corresponds to two atomic monolayers.
The essential parameter of the model that has to be
defined is the effective molecular field $H_z^{\rm eff}$ acting on impurity
spins. One may suppose that it should exponentially decay in the depth
of the barrier. We have, therefore, set it to
$\mu_B H_z^{\rm eff}=5$~meV\ (58~K) that
is of two order less than the critical temperature in the bulk Co.

\begin{figure}[t]
\begin{center}
\includegraphics[scale=0.5, angle=-90]{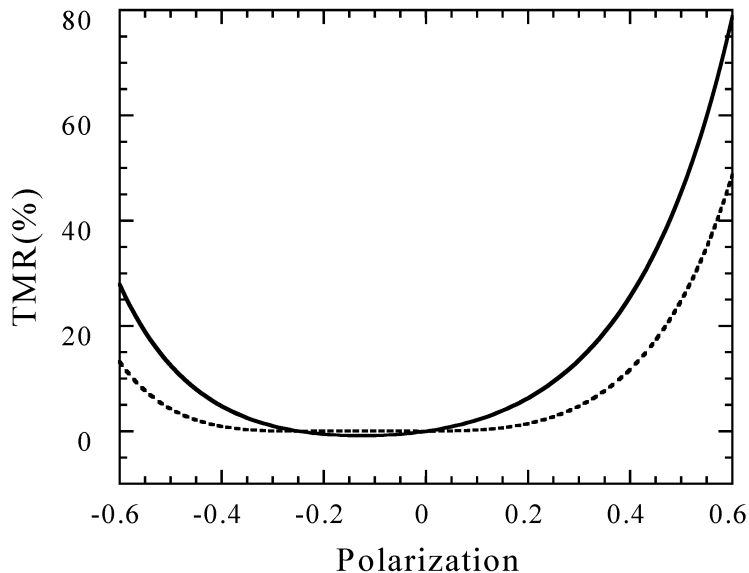}
\end{center}
\caption{\small Tunnel magnetoresistance at $T=0$ as a function of polarization
$P=\frac{\displaystyle k^F_{\uparrow}-k^F_{\downarrow}}
{\displaystyle k^F_{\uparrow}+k^F_{\downarrow}}$ under fixed 
$k^F_{\uparrow}=1.09~\AA^{-1}$.
Other parameters are: $q_0=0.56~\AA^{-1}$, $m_0=0.4$. Solid line corresponds to 
the case of impurity concentration 
$x=8\times 10^{-5}$, dashed line represents the case of absence of impurities.}
\end{figure}

   Consider, first, the case $T=0$. It is possible to estimate the 
concentration
of impurity atoms so that its contribution to the resonance conductivity is
comparable with the ordinary tunnel conductance of the system. One can write
that the impurity density of states (see exp.(\ref{s_Larkin}))
$\nu(\varepsilon_F)=\frac{\displaystyle N_i}
{\displaystyle S w_1 (\Delta\varepsilon)}$, where $N_i$ is a total number of
impurities and $\Delta\varepsilon$
is the width of its energy distribution. On the
other hand $N_i=x N$ and $N=S w_1/a_0^3$ where $N$ is
the total number of atoms in the layer which contains the impurities,
$x$ is the local concentration of impurities in this
volume and $a_0$ is the lattice constant. This yields
$\nu(\varepsilon_F)= x\left(\frac{\displaystyle w_1}
{\displaystyle a_0}\right)\frac{\displaystyle 1}{\displaystyle
a_0^2\Delta\varepsilon}$.
We introduce the characteristic concentration $x_0$
defined so that in the case of
nonmagnetic impurity, the impurity conductance~(\ref{s_Larkin})
is equal to the tunnel conductance (\ref{s0_approx}) of spin $\uparrow$ channel 
in the parallel magnetic configuration
of the ferromagnetic layers. Such definition leads to
\begin{equation}
x_0 \simeq \frac{\Delta\varepsilon}{\pi^2}\left(\frac{2 m_0
a_0^2}{\hbar^2}\right)
\left(\frac{w}{w-w_1}\right)\left(\frac{a_0}{w_1}\right)
\frac{k^F_{\uparrow} q_0 m_0}
{{k^F_{\uparrow}}^2 m_0^2 + q_0^2}\frac{(q_0 w_1)\exp{(-2 q_0 w_1)}}
{1-\exp{(-2 q_0 w_1)}}
\label{x0}
\end{equation}
If we choose $\Delta\varepsilon=0.2$~eV,
then $x_0 = 6.5\times 10^{-5}$. The
conductance of the system at $T=0$ can be extract
from the general expression (14).
We suppose that for both parallel and antiparallel configurations the left
electrode has "up" magnetization and, hence,
$H_z^{\rm eff}$ is positive in both cases.
At zero temperature, all spin-flip processes are frozen and due to the
above assumption, only the configuration of impurity spin with $m_s=1$ is
possible. As a result, only two resonance channels from many
possible ones have nonzero contribution to conductivity, namely
$|\uparrow, m_s=1\rangle \leftrightarrow |\uparrow, m_s=1\rangle$ and
$|\downarrow, m_s=1\rangle \leftrightarrow |\downarrow, m_s=1\rangle$ with
$m_j=3/2$ and $1/2$ respectively. From exp.~(A1) (see Appendix~A)
it follows that
the channel with $m_j = 3/2$ gives
the main contribution into the conductivity
at low temperatures and
${}^P\sigma^{\uparrow\uparrow}_{3/2}
\sim \Gamma_{\uparrow}$ and
${}^{AP}\sigma^{\uparrow\uparrow}_{3/2} \sim \Gamma_{\downarrow}$.
So these contributions depend on the mutual orientation
of the magnetization  of the ferromagnetic layers, therefore
they increase the total amplitude of the TMR.
The total expression for
$TMR=\frac{\displaystyle \sigma^P - \sigma^{AP}}
{\displaystyle \sigma^{AP}}$ including all possible
channels
may be written as
\begin{equation}
TMR = \frac{(\Gamma_{\uparrow}-\Gamma_{\downarrow})(\Gamma_{\uparrow} -
\Gamma_{\downarrow}
 + \frac{x}{x_0}\Gamma_{\uparrow}
 (1-\frac{1}{9}\Gamma_{\downarrow}/\gamma_{\uparrow}))}
 {\Gamma_{\uparrow}\Gamma_{\downarrow}(2 +
 \frac{x}{x_0} (1+\frac{1}{9}\Gamma_{\uparrow}/\gamma_{\uparrow}))}
\end{equation}
where $x$ is concentration, $x_0$ is defined
by~(\ref{x0}), $\gamma_{\uparrow} = \frac{2}{3}\Gamma_{\uparrow} +
\frac{1}{3}\Gamma_{\downarrow}$ and
$\Gamma_{\uparrow(\downarrow)} = k^F_{\uparrow(\downarrow)}q_0^2/
(k^{F2}_{\uparrow(\downarrow)} m_0^2+q_0^2)$  are the tunneling
density of states for $\uparrow(\downarrow)$ spin electrons.
The dependence of TMR effect versus the polarization $P=(k^F_{\uparrow}-
k^F_{\downarrow})/(k^F_{\uparrow}+k^F_{\downarrow})$ is shown in Fig.~3
in comparison with the non-resonant
tunnel conductance at $x=0$.
The total TMR amplitude is larger than the TMR due to direct
tunneling, in accordance with considerations written above.
For a given polarization $P=0.44$ in case of chosen parameters,
the contribution of the impurity assisted tunneling leads to strong
enhancement of TMR amplitude (typically by a factor 2, see Fig.~3).

  We note, that in the case of nonmagnetic impurities, distributed
in the vicinity of only one contact, the resonant impurity conductance
$\sigma^{\rm imp} \sim \Gamma_{\uparrow} + \Gamma_{\downarrow}$ is equal
for both parallel and antiparallel configurations and, therefore,
in this case the mechanism of impurity assisted tunneling is not able
to enhance the TMR effect. The enhancement of TMR amplitude in the
case of paramagnetic impurities is essentially due to the presence
of ferromagnetic exchange coupling between the magnetization
in ferromagnetic electrode and the impurity spins which tends
to induce a ferromagnetic order in the plane of impurities and,
as the result, leads to the preference of impurity spin to be found
in the quantum state with $m_s = 1$.

\begin{figure}
\begin{center}
\includegraphics[scale=0.5, angle=-90]{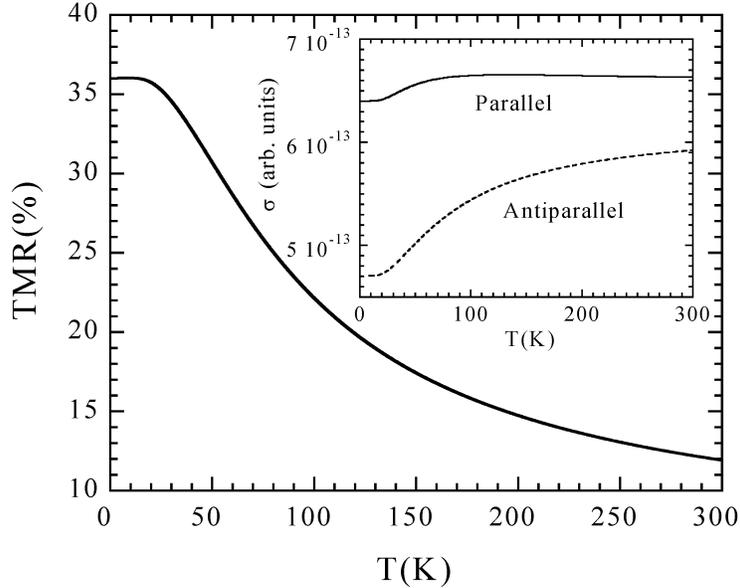}
\end{center}
\caption{\small Tunnel magnetoresistance as a function of the temperature.
(On the insertion: conductance $\sigma$ for the parallel and antiparallel 
configurations.)
Parameters are: $k^F_{\uparrow}=1.09~\AA^{-1}$, $k^F_{\downarrow}=0.42~\AA^{-
1}$,
$q_0=0.56~\AA^{-1}$, $m_0=0.56$. Concentration of impurities $x=8\times 10^{-
5}$.}
\end{figure}

\begin{figure}
\begin{center}
\includegraphics[scale=0.75, angle=0]{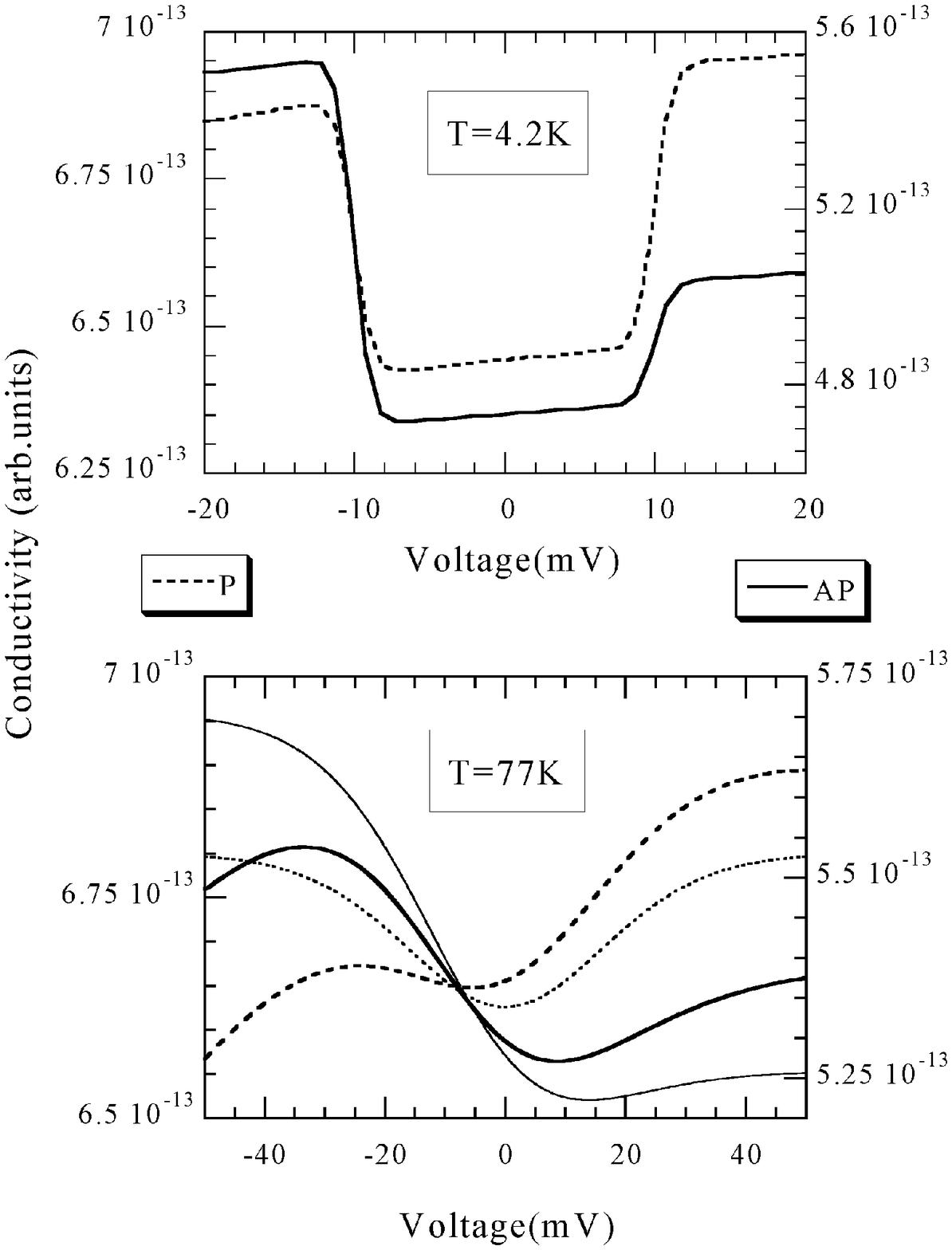}
\end{center}
\caption{\small Differential conductance as a function of the bias voltage at 
$T=4.2~K$ and $77~K$.
Parameters are: $k^F_{\uparrow}=1.09~\AA^{-1}$, $k^F_{\downarrow}=0.42~\AA^{-
1}$,
$q_0=0.56~\AA^{-1}$, $m_0=0.4$. Thick dashed and solid lines correspond to the
conductance in the parallel and antiparallel configurations respectively, 
calculated in the
WKB approximation. For the comparison, the same dependences at $T=77~K$, 
calculated by
approximate formulae, are indicated by thin lines. Concentration of impurities 
$x=8\times 10^{-5}$.}
\end{figure}

  The temperature dependences of resonant conductances for parallel and
antiparallel configurations in the interval from 4.2~K~-- 300~K are presented
in Fig.~4. In the case of parallel alignment,
$\sigma_{\rm imp}^{P}(T)$ is nearly independent on the temperature, but in
antiparallel situation there is a 50\% increase of impurity conductance
$\sigma_{\rm imp}^{AP}(T)$. This originates from
the thermal excitation of both spin-flip and
spin-conserving processes which are frozen at zero temperature.
For AP configuration the process
$|\uparrow, m_s=0 \rangle \to |\downarrow, m_s=1\rangle$
was forbidden at $0^{\circ}$K but now it is allowed
and gives a large contribution into the current as
it is proportional to the product of the largest
density of states $\Gamma_{\uparrow}\Gamma_{\uparrow}$.
As a consequence, the TMR effect
decreases with the increase of the temperature.

   We have also calculated the dependence of the differential conductances on 
bias voltage according to~(\ref{I_VT})
and~(\ref{j_imp}). These dependences at $T=4.2$~K and $T=77$~K are
presented in Fig.~5. A new effect is predicted: the voltage dependence of
the conductance in the antiparallel alignment of magnetization in
ferromagnetic layers is asymmetric under forward and inverse bias voltage
when the paramagnetic impurities inside the insulator layer are
distributed close to only one of the interfaces and are bound by exchange
interaction with magnetization of the nearest ferromagnetic layer.

One can distinguish two different mechanisms that give rise to the
presented asymmetrical behavior with respect to inversion of bias
voltage. The first one, which we refer as zero bias anomaly,
manifests itself at low voltages of the order of 10~mV
(for particular chosen parameters in our model) and is strongly pronounced
only at low temperatures (see Fig.~5, the case of $T=4.2$~K).
It originates from the excitations of spin-flip processes on
the impurity centers. One may look at the general expression
(\ref{I_VT}) and consider the case of low temperature.
An electron undergoing spin-flip scattering, may transfer an amount of energy
$\omega_0=2\mu_B H^{\rm eff}_z$ to the impurity spin thus exiting it at
an higher energy level or on the contrary may acquire this quantum of energy
from it. The latter process is impossible at low temperature.
The former one is possible only if an electron moving, say, from the left
contact possesses an excess energy of at least
$\omega_0$ with respect to Fermi level in the right contact.
The only one process that contributes to
this anomaly at low temperature is
$\phi_{1/2}^{\downarrow}\rightarrow\phi_{1/2}^{\uparrow}$
(see exp.~\ref{phi}). For antiparallel alignment of the magnetization
its quantum mechanical probability
is proportional to ${}^{AP}\sigma_{1/2}^{\downarrow\uparrow}\sim\frac29
\Gamma_{\downarrow}^2/\gamma_{\uparrow}$ for electrons moving from the left
ferromagnetic layer into the right one and is proportional to
${}^{AP}\sigma_{1/2}^{\uparrow\downarrow}
\sim\frac29\Gamma_{\uparrow}^2/\gamma_{\uparrow}$
in the case of electrons moving from the right to the left.
For parallel configuration of magnetizations these probabilities
are equal in both directions and are proportional to
${}^P\sigma_{1/2}^{\downarrow\uparrow}=
{}^P\sigma_{1/2}^{\uparrow\downarrow}
\sim\frac29 \Gamma_{\downarrow}\Gamma_{\uparrow}/\gamma_{\uparrow}$.
As a result, zero bias anomaly at $T = 4.2$~K looks asymmetrically
in the case of antiparralel configuration and is symmetrical in the
case of parallel alignment of magnetizations.

  The differential conductances as a functions of the bias voltage
at $T=77$~K have been calculated using two different
approximations. Thick dashed and solid lines correspond to the
conductances in the parallel and antiparallel configurations,
respectively, that have been calculated by means of WKB
approximation in accordance with expressions
(\ref{I_V}) and (\ref{j_imp}). For the sake of comparison,
the same dependences, indicated by thin lines, have been
calculated with the use approximate formulae (\ref{I_VT}),
where the dependence of $\hat t$--matrix elements on the applied
voltage has been neglected. In course of this, the latter
curves demonstrate the only zero bias anomaly discussed above,
which is substantially smoothed, compared with the case
of $T=4.2$~K. On the contrary, the WKB scheme of calculation
takes into account the variation of the potential profile
inside the insulating barrier under applied bias voltage.
In view of this, the differential conductances calculated by
this scheme exhibit the tendency to increase at the direct
bias voltage and to decrease at the reverse one. As it was shown
above (see section 2.4), this behavior originates
from the shift of the resonant levels inside the insulator due
to externally applied electric field. This second mechanism in the
origin of non-linear voltage dependence of impurity assisted
conductance does not relate with the excitation of spin-flip
processes. It becomes apparent at the voltages of the order
of 50~mV and leads to the asymmetric voltage bias  dependences
in both cases of parallel and antiparallel configurations.

\begin{figure}
\begin{center}
\includegraphics[scale=0.5, angle=-90]{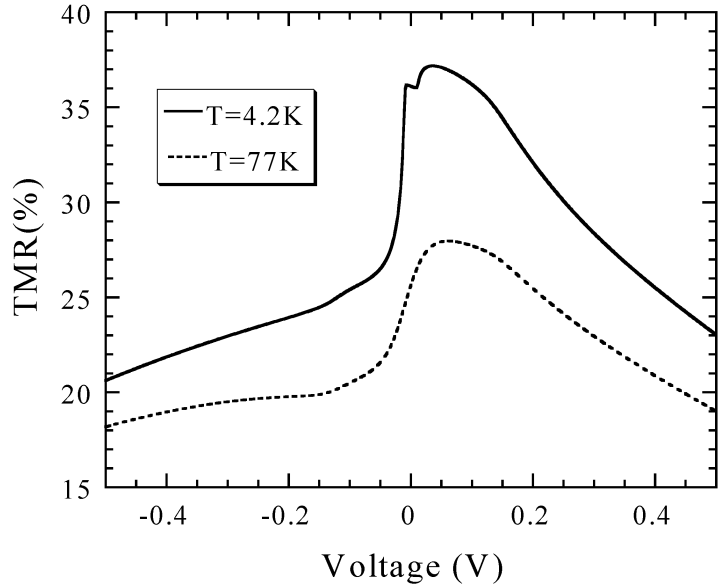}
\end{center}
\caption{\small Tunnel magnetoresistance as a function of the bias voltage: 
solid line - 
$T=4.2~K$, dashed line - $T=77~K$. Parameters are: $k^F_{\uparrow}=1.09~\AA^{-
1}$,
$k^F_{\downarrow}=0.42~\AA^{-1}$, $q_0=0.56~\AA^{-1}$, $m_0=0.4$. 
Concentration of impurities $x=8\times 10^{-5}$.}
\end{figure}

  Finally, the TMR amplitude as a function of the bias voltage is shown
in Fig.~6 for the broad range of applied voltage. Its non-linear
and asymmetric behavior in the range of 0.2~V originates from
the asymmetry of the shifts of resonant impurity levels with
respect to forward and inverse bias. The low bias voltage
anomaly at 10~mV is also strongly pronounced at the curve
corresponding to 4.2~K. The relative contribution of the impurity
assisted conductance to the total current of electrons is
diminished after the value of applied bias voltage exceeds the
half width of impurity band $\Delta\varepsilon/2 = 0.1$~eV.  Therefore,
the TMR amplitude drops to value $\simeq 20\%$ at $0.5$~V
corresponding primary to the pure tunnel conductance.

\section*{Acknowledgments}

A.~Vedyayev and D.~Bagrets are grateful to
CEA/\-Grenoble/\-DRFMC/\-SP2M/\-NM for hospitality.
This work was partially supported by Russian Foundation
for Basic Research (grant No.~98--02--16806).

\section*{Appendix A}
  Let $w_1=c-a$ and $w_2=b-c$ be the position of impurity with respect
to the left and right interfaces, respectively,  and $w=b-a$ be the width
of the insulator layer. We introduce  tunneling densities of states for
spin $\uparrow(\downarrow)$ electrons
$\Gamma_{\uparrow(\downarrow)}~=~k^F_{\uparrow(\downarrow)}q_0^2/
(k^F_{\uparrow(\downarrow)}m_0^2 + q_0^2)$ and denote
$\gamma_{\uparrow}=\frac23\Gamma_{\uparrow}+\frac13\Gamma_{\downarrow}$,
$\gamma_{\downarrow}=\frac13\Gamma_{\uparrow}+\frac23\Gamma_{\downarrow}$.
Then, the position-dependent quantum mechanical probabilities
$\sigma_{m_j}^{\mu\rho}(c)$ can be found as follows.

(a) In case of parallel configuration:

\[
\sigma^{\uparrow\uparrow(\downarrow\downarrow)}_{\frac32(-\frac32)}(c) =
\frac{ \Gamma_{\uparrow(\downarrow)}^2 e^{-2 q_0 w} {\scriptstyle /(w_1 w_2)} }
{\Gamma_{\uparrow(\downarrow)}\frac{\displaystyle e^{ -2 q_0 w_1}}{w_1} +
\Gamma_{\uparrow(\downarrow)}
\frac{\displaystyle e^{- 2 q_0 w_2}}{w_2} },
\]

\[
\sigma^{\uparrow\uparrow(\downarrow\downarrow)}
_{\frac{1}{2}(-\frac{1}{2})}(c)
= \frac{4}{9}
\frac{\Gamma_{\uparrow(\downarrow)}^2 e^{-2 q_0 w} {\scriptstyle /(w_1 w_2)} }
{\gamma_{\uparrow(\downarrow)}\frac{\displaystyle e^{- 2 q_0 w_1}}{w_1} +
\gamma_{\uparrow(\downarrow)}
\frac{\displaystyle e^{-2 q_0 w_2}}{w_2}},
\]

$$
\sigma^{\uparrow\uparrow(\downarrow\downarrow)}
_{-\frac{1}{2}(\frac{1}{2})}(c)
= \frac{1}{9}
\frac{\Gamma_{\uparrow(\downarrow)}^2 e^{-2 q_0 w} {\scriptstyle /(w_1 w_2)} }
{\gamma_{\downarrow(\uparrow)}\frac{\displaystyle e^{- 2 q_0 w_1}}{w_1} +
\gamma_{\downarrow(\uparrow)}
\frac{\displaystyle e^{-2 q_0 w_2}}{w_2}}, \eqno(A1)
$$

\[
\sigma^{\downarrow\uparrow}_{\frac{1}{2}}(c) =
\sigma^{\uparrow\downarrow}_{\frac{1}{2}}(c)
= \frac{2}{9}
\frac{\Gamma_{\uparrow}\Gamma_{\downarrow} e^{-2 q_0 w}
 {\scriptstyle /(w_1 w_2)} }
{\gamma_{\uparrow}\frac{\displaystyle e^{- 2 q_0 w_1}}{w_1} + \gamma_{\uparrow}
\frac{\displaystyle e^{-2 q_0 w_2}}{w_2}},
\]

\[
\sigma^{\uparrow\downarrow}_{-\frac{1}{2}}(c) =
\sigma^{\downarrow\uparrow}_{-\frac{1}{2}}(c)
= \frac{2}{9}
\frac{\Gamma_{\uparrow}\Gamma_{\downarrow} e^{-2 q_0 w}
{\scriptstyle /(w_1 w_2)} }
{\gamma_{\downarrow}\frac{\displaystyle e^{- 2 q_0 w_1}}{w_1} +
\gamma_{\downarrow}
\frac{\displaystyle e^{-2 q_0 w_2}}{w_2}};
\]

(b) In case of antiparallel configuration:
\[
\sigma^{\uparrow\uparrow(\downarrow\downarrow)}_{\frac32(-\frac32)}(c) =
\frac{ \Gamma_{\uparrow}\Gamma_{\downarrow} e^{-2 q_0 w}
{\scriptstyle /(w_1 w_2)} }
{\Gamma_{\uparrow(\downarrow)}\frac{\displaystyle e^{ -2 q_0 w_1}}{w_1} +
\Gamma_{\downarrow(\uparrow)}
\frac{\displaystyle e^{- 2 q_0 w_2}}{w_2} },
\]

\[
\sigma^{\uparrow\uparrow(\downarrow\downarrow)}
_{\frac{1}{2}(-\frac{1}{2})}(z)
= \frac{4}{9}
\frac{\Gamma_{\uparrow}{\Gamma_{\downarrow}} e^{-2 q_0 w}
{\scriptstyle /(w_1 w_2)} }
{\gamma_{\uparrow(\downarrow)}\frac{\displaystyle e^{- 2 q_0 w_1}}{w_1} +
\gamma_{\downarrow(\uparrow)}
\frac{\displaystyle e^{-2 q_0 w_2}}{w_2}},
\]

$$
\sigma^{\uparrow\uparrow(\downarrow\downarrow)}
_{-\frac{1}{2}(\frac{1}{2})}(z)
= \frac{1}{9}
\frac{\Gamma_{\uparrow}{\Gamma_{\downarrow}} e^{-2 q_0 w}
 {\scriptstyle /(w_1 w_2)} }
{\gamma_{\downarrow(\uparrow)}\frac{\displaystyle e^{- 2 q_0 w_1}}{w_1} +
\gamma_{\uparrow(\downarrow)}
\frac{\displaystyle e^{-2 q_0 w_2}}{w_2}},
\eqno(A2)
$$

\[
\sigma^{\uparrow\downarrow(\downarrow\uparrow)}_{\frac{1}{2}}(z)
= \frac{2}{9}
\frac{\Gamma_{\uparrow(\downarrow)}^2 e^{-2 q_0 w} {\scriptstyle /(w_1 w_2)} }
{\gamma_{\uparrow}\frac{\displaystyle e^{- 2 q_0 w_1}}{w_1} +
\gamma_{\downarrow}
\frac{\displaystyle e^{-2 q_0 w_2}}{w_2}},
\]

\[
\sigma^{\uparrow\downarrow(\downarrow\uparrow)}_{-\frac{1}{2}}(z)
= \frac{2}{9}
\frac{\Gamma_{\uparrow(\downarrow)}^2 e^{-2 q_0 w} {\scriptstyle /(w_1 w_2)} }
{\gamma_{\downarrow}\frac{\displaystyle e^{- 2 q_0 w_1}}{w_1} +
\gamma_{\uparrow}
\frac{\displaystyle e^{-2 q_0 w_2}}{w_2}}.
\]

The statistical probabilities $P^{\mu\rho}_{j}(h)$ are independent of the
configuration of the system. We denote $h=\mu_B H^{\rm eff}_z/kT$ and
$Z=2 \cosh{(2h)} +1$, then
\[
P_{\frac32}^{\uparrow\uparrow}(h)=P_{\frac12}^
{\downarrow\downarrow}(h)=Z^{-1} e^{2h}, \qquad
P_{\frac12}^{\uparrow\uparrow}(h)=
P_{-\frac12}^{\downarrow\downarrow}(h)= Z^{-1},
\]
$$
P_{-\frac12}^{\uparrow\uparrow}(h)=P_{-\frac32}^
{\downarrow\downarrow}(h)=Z^{-1} e^{-2h},
\eqno(A3)
$$
\[
P_{\frac12}^{\uparrow\downarrow}(h)=P_{\frac12}^
{\downarrow\uparrow}(h)=Z^{-1}\frac{h e^{h}}{\sinh{(h)}},
\qquad P_{-\frac12}^{\uparrow\downarrow}(h)=P_{-\frac12}
^{\downarrow\uparrow}(h)=Z^{-1}\frac{h e^{-h}}{\sinh{(h)}}
\]

\section*{Appendix B}
The non-trivial functions $I^{\mu\rho}_{m_j}$ are written as follows
\begin{eqnarray}
I^{\downarrow\uparrow}_{1/2}(V, H^{\rm eff}_z)&=&\frac{(eV-2\mu_B
H^{\rm eff}_z)(e^{eV/kT}-1)}
{(e^{(eV-2\mu_B H^{\rm eff}_z)/kT}-1)(2\cosh(2\mu_B H^{\rm eff}_z/kT)+1)},
\nonumber \\
I^{\uparrow\downarrow}_{1/2}(V, H^{\rm eff}_z)&=&\frac{(eV+2\mu_B
H^{\rm eff}_z)(e^{eV/kT}-1)e^{2\mu_B H/kT}}
{(e^{(eV+2\mu_B H^{\rm eff}_z)/kT}-1)(2\cosh(2\mu_B H^{\rm eff}_z/kT)+1)},
\nonumber \\
I^{\downarrow\uparrow}_{-1/2}(V, H^{\rm eff}_z)&=&\frac{(eV+2\mu_B
H^{\rm eff}_z)(e^{eV/kT}-1)}
{(e^{(eV+2\mu_B H^{\rm eff}_z)/kT}-1)(2\cosh(2\mu_B H^{\rm eff}_z/kT)+1)},
\nonumber\\
I^{\uparrow\downarrow}_{-1/2}(V, H^{\rm eff}_z)&=&\frac{(eV-2\mu_B
H^{\rm eff}_z)(e^{eV/kT}-1)e^{-2\mu_B H/kT}}
{(e^{(eV-2\mu_B H^{\rm eff}_z)/kT}-1)(2\cosh(2\mu_B H^{\rm eff}_z/kT)+1)}.
\nonumber
\end{eqnarray}
All the other ones, not written above, are equal to $eV$.

\small{

}
\end{document}